\documentclass[a4paper]{jpconf}

\usepackage{graphicx}
\usepackage{epsfig}
\usepackage{epstopdf}
\usepackage{rotating}
\usepackage[mathscr]{eucal}
\usepackage{amsmath,amssymb,bm,graphics}
\usepackage[pdftex,letterpaper=true]{hyperref}
\usepackage{yfonts}
\usepackage{slashed}

\newcommand\p{k}
\newcommand\omg{\gamma}

\begin{document}

\title
{Ratios of Fluctuation Observables in the Search for the QCD Critical Point \footnote[1]{Talk given by C. Athanasiou at Hot Quarks 2010.}}

\author{Christiana~Athanasiou$^a$, Krishna~Rajagopal$^a$ and Misha~Stephanov$^b$}
\address{$^a$Department of Physics, Massachusetts Institute of Technology, Cambridge, MA 02139, USA $^b$Department of Physics, University of Illinois, Chicago, Illinois 60607, USA \\[1ex] MIT-CTP/4174}

\begin{abstract} 
The QCD critical point can be found in heavy ion collision experiments via the non-monotonic behavior of many fluctuation observables as a function of the collision energy.  The event-by-event fluctuations of various particle multiplicities %and momenta 
are enhanced in those collisions that freeze out near the critical point.  Higher, non-Gaussian, moments of the event-by-event distributions of such observables are particularly sensitive to critical fluctuations, since their magnitude depends on the critical correlation length to a high power.  We present quantitative estimates of the contribution of critical fluctuations to the third and fourth moments of the pion and proton, 
as well as estimates of various measures of pion-proton correlations, all as a function of the same five non-universal parameters.  We show how to use nontrivial but parameter independent ratios among these more than a dozen fluctuation observables to discover the critical point. We also construct ratios that, if the critical point is found, can be used to overconstrain the values of the non-universal parameters.  
\end{abstract}

\section{Introduction}

One of the distinctive features of the phase diagram is the critical point at which the first-order transition between hadron matter and QGP ends. We currently do not have a systematic way of locating this point from first principles as model and lattice calculations face many challenges and much work still needs to be done in order to overcome them.  (For review, see e.g.,~\cite{Stephanov:2004wx}.)  
In the meantime, if the critical point is located in a region accessible to heavy-ion collision experiments, it can be discovered experimentally in current and future experiments at RHIC, BNL and SPS. It is therefore important to define and select
experimental observables that will allow us to locate the critical point, if it is located in an experimentally accessible region.

Upon scanning in the center of mass energy $\sqrt{s}$ and thus in $\mu_B$, one should then be able to locate (or rule out the presence of) the critical point 
by using observables that are sensitive to the proximity of the freeze-out point to the critical 
point~\cite{Stephanov:1998dy,Stephanov:1999zu}, such as pion and proton multiplicity fluctuations. As we vary $\sqrt{s}$, if the freeze-out point approaches the critical point, we would see an increase and then a decrease, as we move away from it, in the fluctuations in the number of these particles. In this paper we describe how to use this non-monotonic behavior in the fluctuations of particle numbers near the critical point as a probe to determine its location.

\section{Results}
\subsection{Cumulant calculations}

The critical mode, $\sigma$, is the mode which develops large long wavelength fluctuations at the critical point. Its screening mass is related to the correlation length $\xi$ by $m_\sigma \equiv \xi^{-1}$. In the case of an infinite system, $\xi$ diverges at the critical point. In reality, $\xi$ reaches a maximum value at the critical point but does not diverge because as it cools the system spends only  a finite time in the vicinity of the critical point. Estimates of the rate of growth of $\xi$ as the collision cools past the critical point suggest that the maximal value of $\xi$ that can be reached is around $1.5-3$ fm \cite{Berdnikov:1999ph,Son:2004iv,Nonaka:2004pg}, compared to the natural $\sim 0.5$ fm away from the critical point.  Near the critical point, the $\sigma^3$ and $\sigma^4$ interaction couplings are given by $\lambda_3 = \widetilde{\lambda}_3 \:T \:(T \:\xi)^{-3/2}, \;\:\:\: \mathrm{and} \:\: \lambda_4 = \widetilde{\lambda}_4  \:(T\: \xi)^{-1},$ where the dimensionless couplings $\widetilde{\lambda}_3$ and $\widetilde{\lambda}_4$ are universal and they have been determined for the Ising universality class~\cite{Tsypin:1994nh}.   Throughout this paper we shall use $\tilde\lambda_3=4$ and $\tilde\lambda_4=12$ as benchmark values, because these are the midpoints of the ranges of values known for these constants~\cite{Tsypin:1994nh,Stephanov:2008qz}.

The long wavelength fluctuations in the $\sigma$-field manifest themselves in observable quantities in so far as
they affect the fluctuations of the occupation numbers of particles that couple to the $\sigma$-field, like protons and pions, which interact with $\sigma$ through the effective Lagrangian: $2 \: G \: \sigma \: \pi^+ \pi^- + g_p\ \sigma\ \bar p\ p$. Throughout this paper we will use $G = 300$ MeV (see Ref.~\cite{Stephanov:1999zu}) and $g_p=7$ 
(see, e.g.,~\cite{Kapusta}) as benchmark values.
It is important to bear in mind that both these parameters and $\tilde\lambda_3$ and $\tilde\lambda_4$ are all uncertain at the  factor of 2 level.

The contribution of critical fluctuations to the 2-, 3- and 4- particle correlators due to $\sigma$-exchanges is given by (see Refs.~\cite{Stephanov:2001zj}, \cite{Stephanov:1999zu} and \cite{Stephanov:2008qz})
\begin{equation} \label{eq:2corr}
 \langle \delta n_\mathbf{\p_1} \delta n_\mathbf{\p_2} \rangle_{\sigma} =  \frac{d^2}{VT}\, g^2\xi^2\frac{v^2_\mathbf{\p_1}}{\omg_\mathbf{\p_1}}  \frac{v^2_\mathbf{\p_2}}{\omg_\mathbf{\p_2}} \ , 
\:\:\:\:\: \langle \delta n_\mathbf{\p_1} \delta n_\mathbf{\p_2} \delta n_\mathbf{\p_3}  \rangle_{\sigma}= 
\frac{2d^3\tilde\lambda_3}{V^2 T^{3/2}} \, g^3 \xi^{9/2} \frac{v^2_\mathbf{\p_1}}{\omg_\mathbf{\p_1}}  \frac{v^2_\mathbf{\p_2}}{\omg_\mathbf{\p_2}}  \frac{v^2_\mathbf{\p_3}}{\omg_\mathbf{\p_3}} ,
\end{equation}
\begin{align} \label{eq:4corr}
 \langle &\delta n_\mathbf{\p_1} \delta n_\mathbf{\p_2} \delta n_\mathbf{\p_3}  \delta  n_\mathbf{\p_4}  \rangle_\sigma=
 \frac{6d^4}{V^3 T^2} \left(2 \tilde\lambda_3^2 - \tilde\lambda_4 \right)g^4 \xi^7\,
 % \notag \\ &\quad\quad \times 
 \frac{v^2_\mathbf{\p_1}}{\omg_\mathbf{\p_1}}  \frac{v^2_\mathbf{\p_2}}{\omg_\mathbf{\p_2}}  \frac{v^2_\mathbf{\p_3}}{\omg_\mathbf{\p_3}}  \frac{v^2_\mathbf{\p_4}}{\omg_\mathbf{\p_4}}\ .
\end{align}
Equations~(\ref{eq:2corr}) - (\ref{eq:4corr}) apply to both protons (with $g=g_p$) and pions (with $g=g_\pi=G/m_\pi$). 
The degeneracy factor $d$ is 2 for both protons and pions since the coupling to the $\sigma$-field is both spin and charge ``blind''.
The variance of the fluctuating occupation number distribution is denoted by $v^2_\mathbf{\p}$ and $\omg_\mathbf{\p} $ is the relativistic gamma-factor of the particle with a given momentum $\mathbf{\p}$.

Mixed pion-proton correlators can also be obtained from (\ref{eq:2corr}) and (\ref{eq:4corr}), with small modifications. For example, the 2 pion - 2 proton correlator is given by
\begin{align} \label{final2p2pi}
 \langle& \delta n_\mathbf{\p_1}^{\pi} \delta n_\mathbf{\p_2}^{\pi} \delta n_\mathbf{\p_3}^{p}  \delta n_\mathbf{\p_4}^{p} \rangle_\sigma=
\frac{6d_\pi^2 d_p^2}{V^3 T^2} \left(2\tilde\lambda_3^2-\tilde\lambda_4\right)g_\pi^2g_p^2 \xi^7 \, %\notag \\&\quad\quad\times 
\frac{v_\mathbf{\p_1}^{\pi\:2}}{\omg_\mathbf{\p_1}^\pi} \frac{v_\mathbf{\p_2}^{\pi\:2}}{\omg_\mathbf{\p_2}^\pi} \frac{v_\mathbf{\p_3}^{p\:2}}{\omg_\mathbf{\p_3}^p} \frac{v_\mathbf{\p_4}^{p\:2}}{\omg_\mathbf{\p_4}^p}.
\end{align}

%%%%%%%%%%%%%%%%%%%%%%%

In order to obtain the critical contribution to the cumulants of the particle multiplicity distributions using the correlators given above, each momentum index needs to be integrated over using $ V\int d^3 \mathbf{\p}/(2\pi)^3.$ A cumulant with $i$ protons and $j$ pions will be denoted by $\kappa_{ipj\pi}$. In order to remove the $V$ dependence of the cumulants we choose to normalize them as
\begin{equation} \label{norm2}
\omega_{ipj\pi}\equiv \frac{\kappa_{ipj\pi}}{\langle N_p \rangle^{i/i+j} \langle N_\pi\rangle^{j/i+j} }\ .
\end{equation}
We can see from expressions (\ref{eq:2corr}) - (\ref{eq:4corr}) that higher cumulants are proportional to higher powers of $\xi$ and thus increase by a larger factor near the critical point where $\xi$ becomes large.

%%%%%%%%

In the case of free particles in the classical Boltzmann regime, with no critical fluctuations, the fluctuations of any particle number obey Poisson statistics.  The Poisson contribution to $\omega_{ip}$ and $\omega_{i\pi}$ is 1, and zero for the mixed cumulants $\omega_{ipj\pi}$.
In reality, this 1 of Poisson statistics gets few percent contributions from Bose-Einstein statistics, from system size fluctuations, from initial state correlations that are incompletely washed out, and from interactions other than those with the fluctuations that are enhanced near the critical point. We are ignoring all of these non-critical corrections to the Poissonian 1. Existing data on the ratio of the fourth to second net proton cumulants,  $\kappa_{4(p-\bar p)}/\kappa_{2(p-\bar p)}$, at $\sqrt{s}=19.6$, 62.4 and 
200 GeV~\cite{Aggarwal:2010wy} confirm that the non-critical corrections to the Poissonian 1 are indeed small.

\subsection{Possible experimental outcomes}

We now illustrate {\em possible} experimental outcomes of measurements of the cumulants,
assuming that the matter produced at the freezeout point of the fireball evolution for some collision energy $\sqrt{s}$ is near the critical point. To start, let us assume that the critical point occurs at $\mu_B^c=400$ MeV and that $\xi_{\rm max}=2$~fm. The simplest ansatz for $\xi(\mu_B)$ that we have found that incorporates the above physics and has roughly the correct scaling behavior from universality of critical phenomena is
\begin{equation}
  \label{eq:xi}
  \xi(\mu_B) = \frac{\xi_{\rm max}}{ \left[ 1 + \frac{(\mu_B-\mu_B^c)^2}{W(\mu_B)^2} \right]^{1/3}}\ ,
\end{equation}
with $W(\mu_B)$ chosen in order to reflect the asymmetry in $\xi(\mu_B)$ when approaching the critical point from the left (smaller $\mu_B$) or from the right. We shall define the width $\Delta$ as the width in $\mu_B$ within which $\xi>1$~fm. We also use our benchmark values of the four nonuniversal parameters that determine the $\omega_{ipj\pi}$ for a given $\xi$, namely
$g_\pi=G/ m_\pi=2.1$, $g_p=7$,  $\widetilde{\lambda}_3=4$ and $\widetilde{\lambda}_4=12$ and we have allowed for the fact that the chemical freeze-out temperature $T$ decreases somewhat with increasing $\mu_B$ by using an empirical parametrization of heavy ion collision data from Ref.~\cite{Cleymans:2005xv}.

%%%%%%%%%%%%%%

\begin{figure}
  \centering
    \includegraphics*[width=0.49\columnwidth]{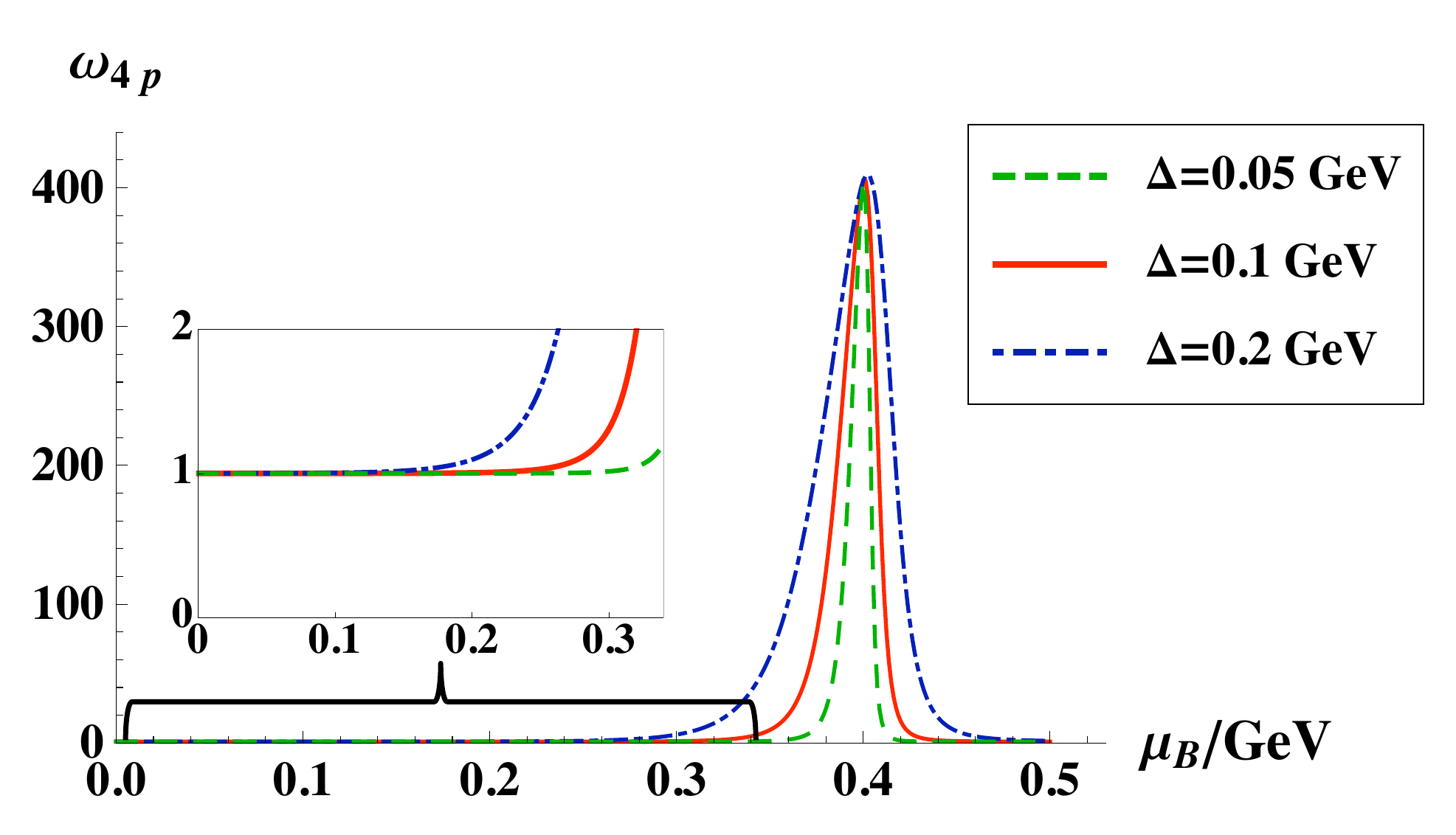}
    \includegraphics*[width=0.49\columnwidth]{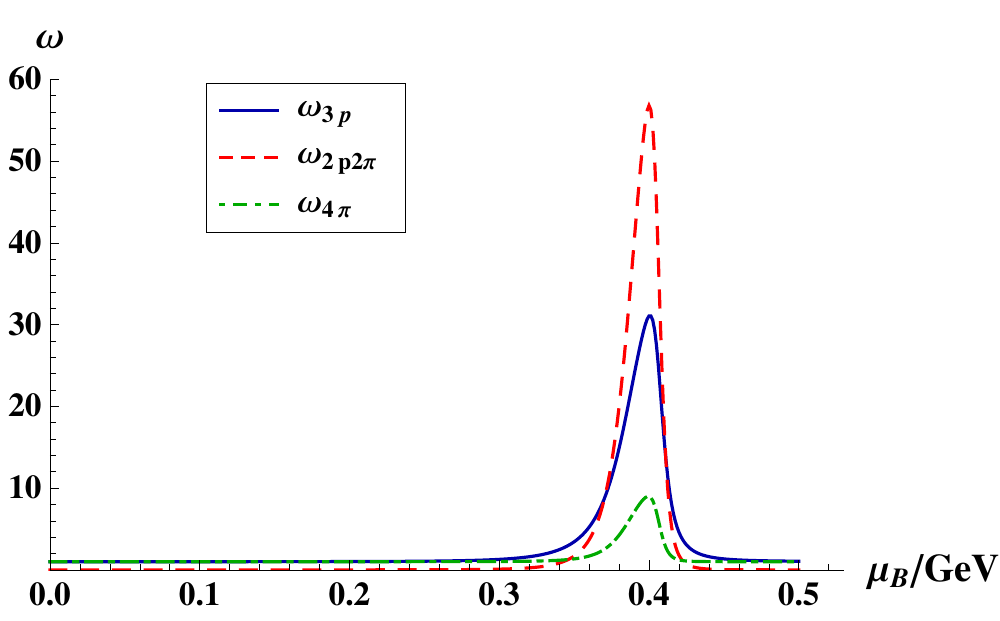}
  \caption{The $\mu_B$-dependence of selected normalized cumulants. We only include the Poisson and critical contributions to the cumulants.  In the left panel we show $\omega_{4p}$ and vary $\Delta$. In the right panel we show other examples of $\omega$'s with $\Delta=100$~MeV. See text for details.
}\label{fig:om4pa}
\end{figure}

Numerical results for some examples of normalized cumulants are shown in Fig.~\ref{fig:om4pa}. We can clearly see the peak in all the normalized cumulants near the critical point. The results indicate that the more protons are involved in the observation measure, the easier it is to identify the critical contribution. 
It is readily apparent that the measurement of these observables in heavy ion collisions at a series of collision energies is very well suited to ruling out (or discovering) the presence of the QCD critical point.

\subsection{Ratios of cumulants}
\label{section3}

In order to locate the critical point, experimental results on multiplicity cumulants will need to be compared to the theoretical predictions of the critical contribution to these cumulants. 
But, recall that we had to choose benchmark values for four parameters: $g_p$, $g_\pi$, $\widetilde{\lambda}_3$, $\widetilde{\lambda}_4$ and also use an ansatz for $\xi(\mu_B)$. 
These parameters are not known reliably or accurately enough to permit a quantitative prediction for the effect of the critical point on any one of the cumulants we have described.   In this Section, we suppose that at some $\sqrt{s}$ there is experimental data showing several of the cumulants significantly exceeding their Poisson values.  We ask how ratios of cumulants can be used to extract information on $\xi$ and the values of the four parameters. And, we construct ratios of cumulants that are independent of $\xi$ and all the parameters, allowing us to make robust predictions for the contribution of critical fluctuations to these ratios. In Table \ref{table:ratios} we present the parameter dependence of various cumulant ratios (with Poisson subtracted before taking the ratios), where we defined $2\tilde{\lambda}_3^2-\tilde{\lambda}_4 \equiv \tilde{\lambda}'_4  $.

%\vspace{5 mm}

\begin{table}  \caption{Parameter dependence of the contribution of critical fluctuations to various particle
    multiplicity cumulant ratios. We have subtracted the Poisson
    contribution from each cumulant before taking the ratio. The table
    shows the power at which the parameters enter in each case.}
\newcommand{\nn}{-}
\begin{tabular}
{cccccccc} \br
ratio &  $g_p$   &   $g_\pi$  & $\tilde{\lambda}_3$ & $ \tilde{\lambda}'_4  $  & $\xi$    \\

	\hline \hline

$\kappa_{2p2\pi}N_\pi/\kappa_{4\pi}\kappa_{2p}$& \nn & $ -2  $ &\nn&\nn&$-2$\\  
\hline
$\kappa_{2p2\pi} N_p^2/\kappa_{4p} N_\pi^2$ &$-2$&2& \nn & \nn & \nn \\
\hline
$\kappa_{3p} N_p^{3/2}/\kappa_{2p}^{9/4}N_\pi^{1/4}$ &$-3/2$& \nn &1& \nn & \nn \\
\hline
$\kappa_{2p}\kappa_{4p}/\kappa_{3p}^2$& \nn & \nn &$-2$&1& \nn \\
\hline
$\kappa_{3p}\kappa_{2\pi}^{3/2}/\kappa_{3\pi} \kappa_{2p}^{3/2}$ & \nn & \nn & \nn & \nn & \nn \\
\hline
$\kappa_{4p}\kappa_{2\pi}^2/\kappa_{4\pi} \kappa_{2p}^2$ & \nn &
\nn & \nn & \nn & \nn \\
\hline
$\kappa_{4p}^3\kappa_{3\pi}^4/\kappa_{4\pi}^3 \kappa_{3p}^4$ & \nn &
\nn & \nn & \nn & \nn \\
\hline
$\kappa_{2p2\pi}^2/\kappa_{4\pi} \kappa_{4p}$ & \nn &
\nn & \nn & \nn & \nn \\
\hline
$\kappa_{2p1\pi}^3/\kappa_{3p}^2 \kappa_{3\pi}$& \nn &
\nn & \nn & \nn & \nn \\
\br
\end{tabular}  \label{table:ratios}
\end{table}

The correlation length $\xi$ and the four nonuniversal parameters always appear in certain
combinations in the multiplicity cumulants and it turns out that we can only
constrain four independent  combinations. The first four rows of Table \ref{table:ratios} show ratios that can be used to constrain one example of four such combinations. The last five entries in the table show
parameter-independent ratios. They have no $\xi$-dependence, no dependence on the four poorly known parameters, and no $\mu_B$-dependence. We find that these five ratios are all precisely 1.

Now let us see how we can use these five ratios in order to locate the
critical point. Suppose that as you change the center of mass energy
$\sqrt s$ of the collisions, experimental evidence for peaks shown in Fig.~\ref{fig:om4pa} begin to emerge. How do you check in a parameter-independent fashion whether the behavior seen in experimental data is consistent with the hypothesis that it is due to critical fluctuations? You first subtract the Poisson contributions, and then construct the last five ratios in Table I.  If the fluctuations seen in this hypothetical data are in fact due to the proximity of the critical point, all five of these ratios will be equal to 1, with no theoretical uncertainties arising from uncertainty in the values of the parameters.  This would be strong evidence indeed for the discovery of the QCD critical point.

\vspace{2mm}

%
%
%
%
%

%\bibliographystyle{utphys}
%\bibliography{references}

\end{document}